\documentclass[prd,aps,amsmath,amssymb,twocolumn]{revtex4}
\usepackage{graphicx}
\usepackage{dcolumn}
\usepackage{bm}

\begin{document}

%

\let\a=\alpha      \let\b=\beta       \let\c=\chi        \let\d=\delta
\let\e=\varepsilon \let\f=\varphi     \let\g=\gamma      \let\h=\eta
\let\k=\kappa      \let\l=\lambda     \let\m=\mu
\let\o=\omega      \let\r=\varrho     \let\s=\sigma
\let\t=\tau        \let\th=\vartheta  \let\y=\upsilon    \let\x=\xi
\let\z=\zeta       \let\io=\iota      \let\vp=\varpi     \let\ro=\rho
\let\ph=\phi       \let\ep=\epsilon   \let\te=\theta
\let\n=\nu
\let\D=\Delta   \let\F=\Phi    \let\G=\Gamma  \let\L=\Lambda
\let\O=\Omega   \let\P=\Pi     \let\Ps=\Psi   \let\Si=\Sigma
\let\Th=\Theta  \let\X=\Xi     \let\Y=\Upsilon

%

%

\def\cA{{\cal A}}                \def\cB{{\cal B}}
\def\cC{{\cal C}}                \def\cD{{\cal D}}
\def\cE{{\cal E}}                \def\cF{{\cal F}}
\def\cG{{\cal G}}                \def\cH{{\cal H}}
\def\cI{{\cal I}}                \def\cJ{{\cal J}}
\def\cK{{\cal K}}                \def\cL{{\cal L}}
\def\cM{{\cal M}}                \def\cN{{\cal N}}
\def\cO{{\cal O}}                \def\cP{{\cal P}}
\def\cQ{{\cal Q}}                \def\cR{{\cal R}}
\def\cS{{\cal S}}                \def\cT{{\cal T}}
\def\cU{{\cal U}}                \def\cV{{\cal V}}
\def\cW{{\cal W}}                \def\cX{{\cal X}}
\def\cY{{\cal Y}}                \def\cZ{{\cal Z}}

\def\dbd{{$0\nu 2\beta\,$}}
%

\newcommand{\Ns}{N\hspace{-4.7mm}\not\hspace{2.7mm}}
\newcommand{\qs}{q\hspace{-3.7mm}\not\hspace{3.4mm}}
\newcommand{\ps}{p\hspace{-3.3mm}\not\hspace{1.2mm}}
\newcommand{\ks}{k\hspace{-3.3mm}\not\hspace{1.2mm}}
\newcommand{\des}{\partial\hspace{-4.mm}\not\hspace{2.5mm}}
\newcommand{\desco}{D\hspace{-4mm}\not\hspace{2mm}}



\title{\boldmath $\rho$ exchange contribution to neutrinoless double beta decay }

\author{Namit Mahajan
}
\email{nmahajan@prl.res.in}
\affiliation{
 Theoretical Physics Division, Physical Research Laboratory, Navrangpura, Ahmedabad
380 009, India
}


\begin{abstract}
We consider $\rho$ meson contributions to neutrinoless double beta decay amplitude stemming from the hadronization of the
short distance quark-electron currents. 
These contributions are evaluated within vacuum dominance approximation. The one and two $\rho$ exchange 
contributions affect the Fermi transition nuclear matrix element
in a way that lead to near cancellations in the same chirality, left-left and right-right, short range amplitudes when these new contributions
are combined with the conventional short range amplitudes, while the left-right amplitude almost triples. This then necessitates the
inclusion of $\rho$ exchange amplitudes in any phenomenological study, like in left-right theories.
\end{abstract}

\maketitle

Experiments have firmly established that the neutrinos, which are massless within the Standard Model (SM) of particle 
physics, have non-zero, albeit tiny mass, and different flavours of neutrinos mix with each other
(see \cite{Esteban:2018azc} for the current best fit values of the parameters). Further, neutrinos
are electrically neutral, allowing them to be their own anti-particles ie can be Majorana in nature \cite{Majorana:1937vz}. 
Neutrinoless double beta decay (\dbd), $(A,Z)\rightarrow (A,Z+2) + 2e^-$, provides an unambiguous way of establishing the Majorana
nature of the neutrinos, and also the lepton number violation \cite{Furry:1939qr}. 
Theoretically as well, \dbd decay is heralded as a useful probe of physics beyond SM,
 having particular relevance for neutrino masses and mass hierarchy. 
For an incomplete list discussing \dbd decay phenomenology and related signatuures see e.g. \cite{Keung:1983uu}.
The search for neutrinoless double beta decay thus constitutes an important endevour. Experimentally, 
studies have been carried out or planned on several nuclei (\cite{KlapdorKleingrothaus:2006ff}-
\cite{Albert:2017hjq}). 
Only one of the experiments \cite{KlapdorKleingrothaus:2006ff}
(HM) has claimed observation of \dbd signal in $^{76}{\mathrm Ge}$. 
The half-life at $68\%$ confidence level is: $T^{0\nu}_{1/2}(^{76}{\mathrm Ge}) 
= 2.23^{+0.44}_{-0.31}\times 10^{25}\,
{\mathrm yr}$. A combination of the first results from Kamland-Zen and EXO-200, both using $^{136}{\mathrm Xe}$, 
yielded a lower limit on the half-life $T^{0\nu}_{1/2}(^{136}{\mathrm Xe}) > 3.4 \times 10^{25}\, {\mathrm yr}$ 
which is at variance with the HM claim, and so is the GERDA result.

Neutrinoless double beta decay process can proceed via the light neutrinos, $\nu_i$'s, (the so called long range part) or due to
 heavy degrees of freedom (the short range part) for example through exchange of heavy neutrinos, $N_i$'s, or other heavy particles in specific
 models like R-parity violating supersymmetric theories or theories with leptoquarks (see \cite{Doi:1985dx} and 
 references therein for a quick review of some of the essential theoretical and experimental issues).
 The short range part due to the heavy physics is due to intermediate particles with masses much larger than the
  relevant scale of the process $\sim {\mathcal{O}}$(GeV), allowing for the heavier degrees 
 of freedom to be systematically integrated out, leaving behind a series of operators built out of low energy fields, 
 the up and  down quarks and electrons,
 weighted by the short distance coefficients, called Wilson coefficients (denoted by $C_A$ below). 
 This provides a very convenient  and systematic framework to evaluate the 
 decay amplitude in terms of short distance coefficients which encode all the information about the high energy physics.
 In this process of integrating out
 the heavy degrees of freedom, the operators and thus the effective Lagrangian obtained is at the typical scale of the 
 heavy particles. Using then the renormalization group equations (RGEs), perturbative QCD effects can be computed. These QCD
 corrections have been shown to be very significant \cite{Mahajan:2013ixa}, in particular due to the
 colour mismatched operators (see also \cite{Gonzalez:2015ady}).
 In this way, the high energy particle physics input gets separated from the low energy dynamics contained in the nuclear
 matrix elements (NMEs) of the quark level operators sandwiched between the nucleon states (see \cite{Simkovic:2007vu}
 to get an idea of different approaches to calculate these NMEs). These NMEs are usually the source
 of large uncertainty to \dbd predictions, and in principle one could compare predictions for various nuclei
 undergoing \dbd transition to understand and eventually reduce the sensitivity on NMEs.

To be concrete, we begin by considering the short range quark-electron operators
(denoted by ${\mathcal{L}}_{\mathrm qe}$ below in the text) and
as an example, consider a heavy right handed neutrino, mass $M_N$, and SM gauge group. The resulting quark level \dbd amplitude takes the form
\begin{eqnarray}
{\mathcal{A}} &\sim&  \underbrace{\frac{(V_{ud} T_{ei})^2}{M_W^4M_N}}_{G}\underbrace{\bar{u}\gamma_{\mu}(1-\gamma_5)d\,
\bar{u}\gamma^{\mu}(1-\gamma_5)d}_{{\mathcal{J}}_{q,\mu}{\mathcal{J}}_q^{\mu}}\, \underbrace{\bar{e}(1+\gamma_5)e^c}_{j_e} \nonumber
\end{eqnarray}
where $V_{ud}$ and $T_{ei}$
are the quark and leptonic mixing matrix elements while $G = G_F^2/M_N$ contains the short distance physics, with $G_F$ being the Fermi constant.
The physical \dbd amplitude is obtained by sandwiching the quark level operators between the initial and final nuclear states, $\vert i\rangle$
and $\vert f\rangle$, finally evaluated in terms of the NMEs:
\begin{equation}
{\mathcal{A}}_{0\nu 2\beta} = \langle f\vert i{\mathcal{H}}_{\mathrm eff}\vert i\rangle \sim G\,
\underbrace{\langle f\vert {\mathcal{J}}_{q,\mu}{\mathcal{J}}_q^{\mu}\vert i\rangle}_{\boldmath NME}\,j_e \nonumber
\end{equation}
The short distance or high energy physics cleanly separates from the low energy matrix elements. 
The low energy effective Lagrangian is expressed as a sum of operators, $O_A$ weighted by the
Wilson coefficents $C_A$: 
${\mathcal{L}}_{\mathrm eff} = G_A C_A O_A$, 
where we have allowed for more than one $G$ for more complicated theories. In the example considered above, there is only one operator 
$O_1 = {\mathcal{J}}_{q,\mu}{\mathcal{J}}_q^{\mu}\,j_e = \bar{u_i}\gamma_{\mu}(1-\gamma_5)d_i\,\bar{u_j}\gamma^{\mu}(1-\gamma_5)d_j\,
\bar{e}(1+\gamma_5)e^c$ ($i,j$ denoting the colour indices) and the corresponding Wilson coefficient $C_1=1$. 
In other models like SUSY with R-parity violation or leptoquarks, Fierz transformations have to be employed to bring the 
operators in form similar to above. The Lorentz and Dirac structure 
of the quark level operator involved decides which NME enters the \dbd rate. Perturbative QCD corrections don't just correct the
Wilson coefficients but also lead to colour mismatched operators (typically with strength that is $1/N_c$ ($N_c=3$ being the number
of colours) of the colour allowed operators) which are then Fierz transformed and can give different operators and thereby bringing
a host of different NMEs which would not have been expected otherwise.

In the usual treatment of the short range part (see \cite{Doi:1985dx}), one considers following simplifications: (i) assume that the final electrons are
emitted in the S-wave ie the long wavelength approximation is employed; (ii) closure approximation ie the energy of the virtual neutrino is much
larger than the nulcear excitation energy, thereby allowing to sum over the intermediate set of states with great ease;
(iii) (non-relativistic) impulse approximation which allows to write the matrix elements of product of the quark currents 
between the nuclear states 
in terms of say the Fermi and Gammow-Teller matrix elements. For the case of quark currents being V-A form, the above procedure
implies that 
\begin{eqnarray}
 {\mathcal{J}}_{q,\mu}(\vec{x_1}){\mathcal{J}}_q^{\mu}(\vec{x_2}) &=& \sum_{n,m}\tau_+^n\tau_+^m\delta(\vec{x_1}-\vec{r_n}) 
 \delta(\vec{x_2}-\vec{r_m}) \nonumber \\
 &&(g_V^2(q^2)-g_A^2(q^2)\vec{\sigma^n}\cdot\vec{\sigma^m})
\end{eqnarray}
which lead to the Fermi (vector part of the current) and Gammow-Teller (axial-vector part of the current) nuclear 
matrix elements, ${\mathcal{M}}_{F}$ and ${\mathcal{M}}_{GT}$ respectively
\begin{equation}
 {\mathcal{M}}_{F} = \langle \Psi_f\vert\sum_{n,m} H(r_n,r_m,\bar{E})\tau_+^n\tau_+^m\vert\Psi_i\rangle
\end{equation}

\begin{equation}
 {\mathcal{M}}_{GT} = \langle \Psi_f\vert\sum_{n,m} H(r_n,r_m,\bar{E})\tau_+^n\tau_+^m \vec{\sigma^n}\cdot\vec{\sigma^m}\vert\Psi_i\rangle
\end{equation}
In the above expressions, $\sigma$ and $\tau$ are Pauli matrices and $\tau_+ = (\tau_1+i\tau_2)/2$; indices $n,m$ run over all the
nucleons in the nucleus and $H(r_n,r_m,\bar{E})$ denotes the heavy neutrino potential. Further, $g_V(0)=1$ and $g_A(0)\simeq 1.27$.
$\Psi_{i,f}$ are the initial and final state nuclear wave-functions. It is
worth mentioning that in writing these matrix elements starting from the quark level currents, the dipole and anapole terms are 
not shown as they are much smaller than those displayed above. For the present purpose, we shall choose to neglect them but they can
be systematically included. See \cite{Simkovic:2007vu} for details and updated numerical values of the NMEs calculated within different schemes.
The differences in the numerical values of NMEs are at the heart of large uncertainties in the predictions.

Considering only vector and axial quark currents, the following set of short distance operators are to be considered (showing only the colour matched
operators; $i,j$ in the operators below denote the colour indices) 
\begin{eqnarray}
O^{LL} &=& \bar{u_i}\gamma_{\mu}(1-\gamma_5)d_i\,\bar{u_j}\gamma^{\mu}(1-\gamma_5)d_j\,\bar{e}(1+\gamma_5)e^c \nonumber \\
O^{RR} &=& \bar{u_i}\gamma_{\mu}(1+\gamma_5)d_i\,\bar{u_j}\gamma^{\mu}(1+\gamma_5)d_j\,\bar{e}(1+\gamma_5)e^c \nonumber \\
O^{LR} &=& \bar{u_i}\gamma_{\mu}(1-\gamma_5)d_i\,\bar{u_j}\gamma^{\mu}(1+\gamma_5)d_j\,\bar{e}(1+\gamma_5)e^c \nonumber \\
O^{RL} &=& \bar{u_i}\gamma_{\mu}(1+\gamma_5)d_i\,\bar{u_j}\gamma^{\mu}(1-\gamma_5)d_j\,\bar{e}(1+\gamma_5)e^c 
\label{lrops}
\end{eqnarray}
with their associated Wilson coefficients, $C^{LL,RR,LR,RL}$. 
These are the set of operators say in left-right theories, with $LR, RL$ stemming from the heavy-light $W$ mixing. From an effective
field theory (EFT) point of view, there will be scalar-pseudoscalar and tensor operators (and their combinations) as well. For definiteness,
we focus on the operators in Eq.(\ref{lrops}). It turns out that the Short Range (SR) NMEs depend on whether the quark chiralities are same or different:
\begin{eqnarray}
 {\mathcal{M}}_{SR}^{LL} = {\mathcal{M}}_{SR}^{RR} &=& g_V^2{\mathcal{M}}_{F} - g_A^2{\mathcal{M}}_{GT} \nonumber \\
 {\mathcal{M}}_{SR}^{LR+RL} &=& g_V^2{\mathcal{M}}_{F} + g_A^2{\mathcal{M}}_{GT}
\end{eqnarray}
It is to be noted that ${\mathcal{M}}_{GT} \sim -(3$-$4)\,{\mathcal{M}}_{F}$ for all the nuclei
and different determinations of NMEs considered (see for example \cite{Simkovic:2007vu}). Below, for simplicity and convenience, we shall assume that the two terms on the right hand
side of the above equations differ by a factor $\sim 5$, after taking into account $g_V$ and $g_A$.
Since the discussion will not be specific to a particular nucleus, we shall assume this for all the nuclei. Therefore, one has the 
approximate results: ${\mathcal{M}}_{SR}^{LL} = {\mathcal{M}}_{SR}^{RR} \sim 6{\mathcal{M}}_{F}$ and ${\mathcal{M}}_{SR}^{LR+RL} 
\sim -4{\mathcal{M}}_{F}$, thereby yielding the following approximate forms for the corresponding contributions to the amplitude:
\begin{eqnarray}
 {\mathcal{A}}_{SR}^{LL/RR} &\sim& 6\,C^{LL/RR}\,{\mathcal{M}}_{F} \nonumber\\
 {\mathcal{A}}_{SR}^{LR/RL} &\sim& -4\,C^{LR/RL}\,{\mathcal{M}}_{F}
 \label{SRamp}
\end{eqnarray}

A given theory of lepton number violation implies a set of quark-lepton level operators comprising the effective
Lagrangian. Using the RGEs and including effects of operator mixing, the final amplitude for the \dbd can be written in terms
of short distance coefficients and various NMEs. The rate thus computed can be then contrasted with the experimental limits on the
half life of the neutrinoless double beta decay process for the specific nucleus and stringent limits are obtained
on the parameters of the theory. Alternatively, an EFT point of view could be adopted and all the relevant operators are then 
written at the hadronic scale, using which the amplitude and thus the decay rate is computed, which then leads to constraints on
different effective coefficients. One of these two is followed in the phenomenological studies.

The above is not the entire story. Till now, the \dbd amplitude is calculated by directly evaluating the nuclear matrix elements from the 
quark currents. However, it was shown in \cite{Faessler:1996ph} that there are additional contributions wherein these quark currents
first hadronise into pions, and then one can evaluate the two pion exchange contributions which were shown to be the dominant ones.
These authors used the on-shell matching conditions to match the matrix elements of quark-lepton currents between the pion states on to
the hadron level effective terms. Ref.\cite{Prezeau:2003xn} provided a systematic EFT set-up to match the quark-electron operators to
a chiral effective theory with two pion and electron ($\pi\pi ee$) vertices as well as two nucleon, one pion and electron ($NN\pi ee$)
vertices and four nucleon and electron ($NNNNee$) vertices, where $\pi$, $N$ and $e$
denote pion, nucleon and electron respectively. It also discussed inclusion of next to leading and next to next to leading order contributions.  
It was confirmed that the two pion exchange contributions indeed dominate though their calculation employed Naive Dimensional Analysis (NDA) 
for the power counting in contrast to the Vacuum Insertion Approximation (VIA) as employed in the on-shell matching in \cite{Faessler:1996ph}.
That VIA may miss some fraction of the total contribution is not an unexpected satatement since in VIA, only the contribution due to
vacuum is retained while that due to other states is neglected. However, VIA provides a simple and quick
approximation to estimate these new effects. 
The one and two pion contributions are non-local as they involve pion propagators. Recall that $m_{\pi^+} \sim 140\,{\mathrm MeV}$
and the typical momentum, $\vec{q}$,
flowing through the propagators is ${\mathcal{O}}(200\,{\mathrm MeV})$. This approach has been further developed and refined in \cite{Cirigliano:2017djv}.
That the two pion contribution can play a rather important role in phenomenological studies has been re-emphasized in \cite{Li:2020flq} in the
context of left-right symmetric theories. It is worth pointing out that almost all phenomenological studies on \dbd do not include these
contributions and as recently pointed out in \cite{Li:2020flq}, the inferences and limits drawn could be significantly altered once these
are included. 

\begin{figure}[ht!]
\vskip 0.32cm
\hskip 1.35cm
\hbox{\hspace{0.03cm}
\hbox{\includegraphics[scale=0.4]{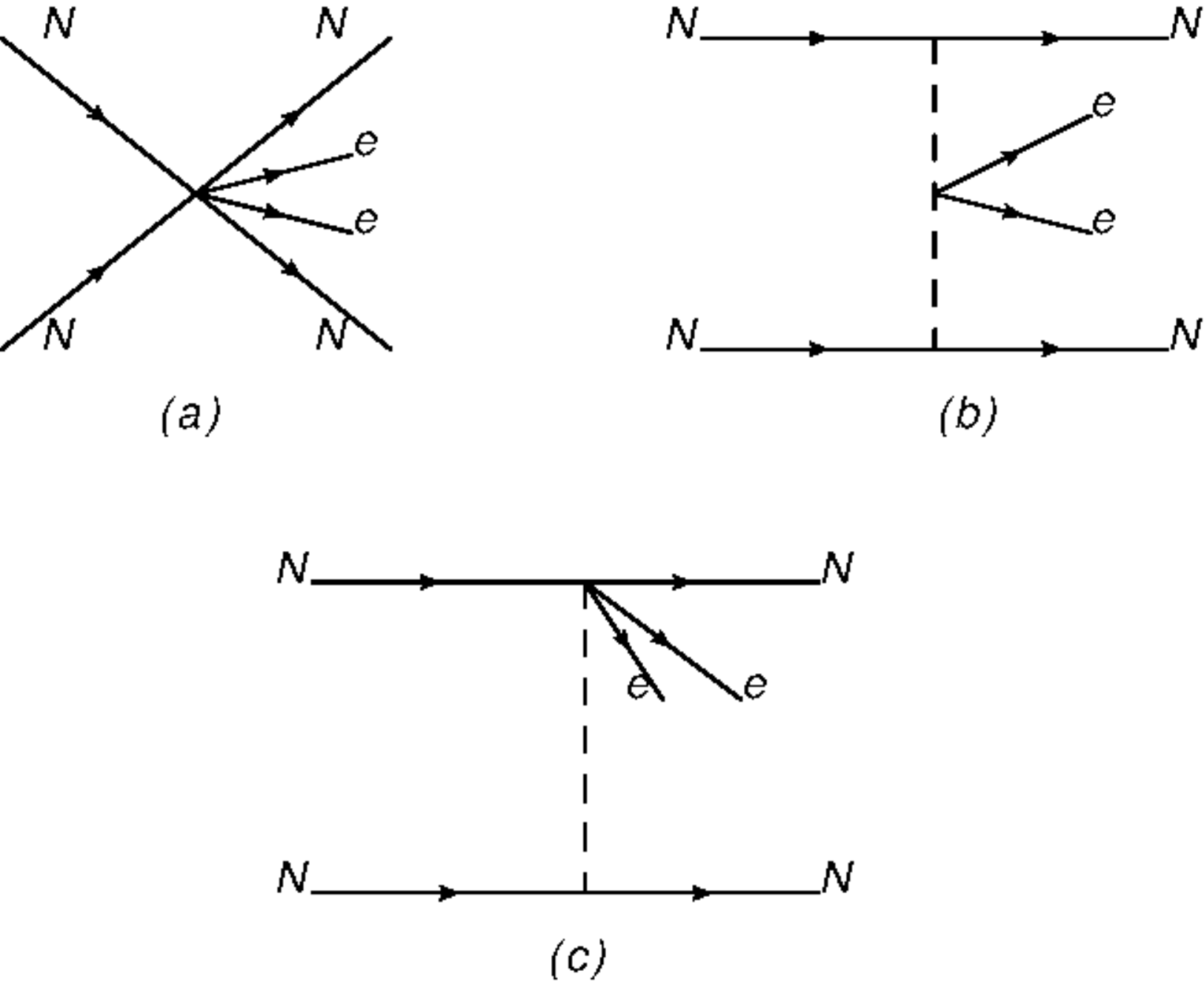}}
}
\caption{Hadronic level diagrams (drawn using JaxoDraw \cite{Binosi:2003yf}) with the intermediate dotted lines denoting a $\rho$ meson. (a)$NNNNee$; (b) Two $\rho$ exchange; 
(c) $NN\rho ee$ at one vertex and other vertex being
$\rho NN$ (possibly including parity violating terms) and the permuted diagram with the vertices exchanged.  
}
\label{fig1}
\end{figure}

Given that there are important contributions to \dbd amplitude beyond those obtained by directly taking the matrix elements of the 
short distance quark currents between the nuclear states emanating from one and two pion exchange diagrams, one is led to ask if there are additional
contributions beyond the pionic ones. In particular, does $\rho$-meson exchange, analogous to pion exchange, yield significant contributions?
We now focus our attention on new contributions coming from hadronization of quark currents into $\rho$ mesons. 
Relevant diagrams are shown in Fig.(\ref{fig1}). Compared to the pion exchange,
there are two differences: (i) $m_{\rho}\, (= 770\, {\mathrm MeV}) >> \vert\vec{q}\vert\, (\sim 200\, {\mathrm MeV})$,
(ii) $m_{\pi} << m_N \,(\sim 940\, {\mathrm MeV})
\sim 1.2m_{\rho}$. Statement (i) means that it is a good approximation to neglect $\vert q\vert$ in comparison with $m_{\rho}$, both in the
numerator and denominator of the $\rho$-meson propagator: $-i(g_{\mu\nu}-q_{\mu}q_{\nu}/m_{\rho}^2)/(q^2-m_{\rho}^2)
\rightarrow ig_{\mu\nu}/m_{\rho}^2$ ie at this level of approximation, the $\rho$ contribution is being treated as a local one. It is to be noted
that this additional local contribution should be distinguished from with theconventional short range one, since having the explicit $\rho$-meson specific
mass and couplings will play an important role, as we see below.

The parity conserving strong $\rho NN$ vertices are described by the phenomenological Lagrangian (see for example \cite{Fischbach:1973vg})
\begin{equation}
 {\mathcal{L}}_{\rho NN} = g_{\rho}\bar{N}\left(\gamma_{\mu} + i\frac{\chi_{\rho}}{2m_N}\sigma_{\mu\nu}q^{\nu}\right)\vec{\tau}\cdot\vec{\rho}^{\mu}N
\end{equation}
where again the nucleon anomalous magnetic moment term, with strength $\chi_{\rho}$ (also denoted as $\mu_N$ in literature), is neglected as
$q << m_N$. There exist various determinations of $g_{\rho}$ and $\chi_{\rho}$ but the combination $g_{\rho}(1+\chi_{\rho})$ takes roughly
the same value $\approx 21$. In what follows, we choose $g_{\rho} \sim 4.5$ (see \cite{Zhu:2004vw} and references therein). Parity violating
terms can also be written. The parity violating couplings are found to be significantly smaller compared to parity conserving ones and therefore,
in the present context such terms are not included.

Consider first the diagram (b) in Fig.(\ref{fig1}).
Analogous to the pion case in \cite{Faessler:1996ph},
we write the relevant term in the hadronic Lagrangian as:
\begin{equation}
 {\mathcal{L}}_{\rho\rho ee} = \frac{G_F^2m_{\rho}^2}{2 m_N}(m_{\rho}^2 a_{2\rho}\vec{\rho}^{\alpha}\cdot\vec{\rho}_{\alpha})
 \,\bar{e}(1+\gamma_5)e^c 
\end{equation}
Using this vertex, along with the parity conserving $\rho NN$ vertex and approximating the $\rho$ propagators as above, 
the amplitude for diagram in Fig. 1(b) takes the form
\begin{equation}
 {\mathcal{A}}_{2\rho} = \frac{G_F^2m_{\rho}^2}{2 m_N} g_{\rho}^2a_{2\rho}[\bar{u_p}\gamma_{\mu}u_n][\bar{u_p}\gamma^{\mu}u_n]\,\bar{e}(1+\gamma_5)e^c
\label{2rhoamp}
 \end{equation}
To obtain $a_{2\rho}$, on-shell matching condition is employed:
\begin{equation}
 \langle\rho,2e\vert{\mathcal{L}}_{qe}\vert\rho\rangle = \langle\rho,2e\vert{\mathcal{L}}_{\rho\rho ee}\vert\rho\rangle
\end{equation}
where to facilitate a quick comparison, the quark-electron Lagrangian is rewritten as
\begin{equation}
 {\mathcal{L}}_{qe} = \frac{G_F^2}{2 m_N} \sum_A\,C^{A}O^{A}\,\bar{e}(1+\gamma_5)e^c
\end{equation}
with $A = LL,\, RR,\, LR+RL$. To complete the on-shell matching, the last step needed is to make use of vacuum, dominance or VIA. This
is achieved through the use of following defining relations (see \cite{Ball:1998sk}):
\begin{eqnarray}
 \langle 0\vert\bar{u}\gamma_{\mu}\vert\rho(P)\rangle &=& f_{\rho}m_{\rho}\epsilon_{\mu} \nonumber \\
  \langle 0\vert\bar{u}\sigma_{\mu\nu}\vert\rho(P)\rangle &=& if^T_{\rho}(\epsilon_{\mu}P_{\nu}-\epsilon_{\nu}P_{\mu})
\end{eqnarray}
with $f_{\rho} = 198\pm 7$ MeV and $f^T_{\rho} = 160\pm 10$ MeV. On-shell matching, assuming VIA, yields 
\begin{equation}
 a_{2\rho} = -\frac{2f_{\rho}^2}{m_{\rho^2}}\sum_A C^A \sim -2.5 \sum_A C^A
\end{equation}
 
 Next, consider the one $\rho$ exchange diagram (c) in Fig.(\ref{fig1}) and its permuted one. The $NN\rho ee$ interaction term is written as
 \begin{equation}
  {\mathcal{L}}_{NN\rho ee} = \frac{G_F^2m_{\rho}^2}{2 m_N}a_{1\rho}\,\bar{N}(\gamma_{\mu}\vec{\tau}\cdot\vec{\rho}^{\mu})N
 \,\bar{e}(1+\gamma_5)e^c 
 \end{equation}
resulting in one $\rho$ exchange contribution to the \dbd amplitude given by
\begin{equation}
 {\mathcal{A}}_{1\rho} = -\frac{G_F^2}{2 m_N} g_{\rho}a_{1\rho}[\bar{u_p}\gamma_{\mu}u_n][\bar{u_p}\gamma^{\mu}u_n]\,\bar{e}(1+\gamma_5)e^c
 \label{1rhoamp}
\end{equation}
Following the same steps as above for on-shell matching and factorizing the quark level product of currents using VIA as: 
$\langle p\vert {\mathcal{J}}_{q,\mu}{\mathcal{J}}_q^{\mu}\vert\rho n\rangle = \langle p\vert {\mathcal{J}}_{q,\mu}\vert n\rangle
\langle 0\vert{\mathcal{J}}_q^{\mu}\vert\rho\rangle$, one obtains after taking into account the combinatoric factors
\begin{equation}
 a_{1\rho} = 4\frac{f_{\rho}g_{\rho}}{m{\rho}} g_V\sum_A C^A \sim 4.5 \sum_A C^A
\end{equation}
where $g_V(0) = 1$ has been used. 

  The inclusion of the $\rho$ exchange diagrams thus produces an extra contribution${\mathcal{A}}^{VIA}_{\rho}$ to \dbd 
\begin{eqnarray}
 {\mathcal{A}}^{VIA}_{\rho} &=& {\mathcal{A}}_{1\rho} + {\mathcal{A}}_{2\rho} \\
&\sim& -\frac{G_F^2}{2 m_N}(7\sum_A C^A)[\bar{u_p}\gamma_{\mu}u_n][\bar{u_p}\gamma^{\mu}u_n]\,\bar{e}(1+\gamma_5)e^c \nonumber
\label{rhoamp}
 \end{eqnarray}
 One immediately notices from the Dirac structure of
 the amplitudes in Eq.(\ref{2rhoamp}) and Eq.(\ref{1rhoamp}) that $\rho$ exchange only results in Fermi transition matrix element.
 The net effect due to ${\mathcal{A}}_{1\rho}$ and ${\mathcal{A}}_{2\rho}$  is to modify the $LL$, $RR$ and $LR/RL$ short range amplitudes, Eq.(\ref{SRamp}):
\begin{eqnarray}
 {\mathcal{A}}_{SR}^{LL/RR} &\rightarrow& 6\,C^{LL/RR}\,{\mathcal{M}}_{F} - 7\,C^{LL/RR} {\mathcal{M}}_{F} \sim 0 \nonumber\\
 {\mathcal{A}}_{SR}^{LR/RL} &\rightarrow& -4\,C^{LR/RL}\,{\mathcal{M}}_{F} - 7\,C^{LR/RL} {\mathcal{M}}_{F} \nonumber \\
 &=& -11\,C^{LR/RL} {\mathcal{M}}_{F}
\end{eqnarray}
This constitutes the main result of this study, namely an almost complete cancellation brought about due to $\rho$ exchange contributions
for the $LL$ and $RR$ short range amplitudes, ${\mathcal{A}}_{SR}^{LL}$ and ${\mathcal{A}}_{SR}^{RR}$, and the mixed left-right
amplitude, ${\mathcal{A}}_{SR}^{LR+LR}$ getting almost tripled. 
These results clearly show that $\rho$ exchange diagrams do impact quite significantly and have the potential to completely
change the phenomenological predictions, and thus call for the phenomenological analyses to be redone. 

In this article, we have studied the $\rho$ exchange contributions to \dbd when the quarks in the short distance product of currents hadronize
into $\rho$ mesons, which are massive compared to the momentum flowing through the internal lines. This allows the $\rho$ propagators to be
shrunk to a point and employing VIA, the one and two $\rho$ meson contributions to neutrinoless double beta decay amplitudes have been estimated.
It is found that for the vector and axial-vector currents, as considered here, the $\rho$ exchange results in a contribution which brings
an almost complete cancellation when combined with the conventional short range left-left and right-right \dbd amplitudes, while the mixed
chirality left-right amplitude gets enhnaced by a factor $\sim 3$. This marks the first step in highlighting the importance of including
$\rho$ exchange diagrams. A similar analysis can be carried for short distance operators with Dirac structure other than the vector/axial-vector
considered in this paper. The two $\rho$ exchange contribution due to tensor currents is quickly seen to lead to very similar results.
Acknowledging the short comings of VIA, a better approach would
be to map the quark-electron operators onto a set of operators in the chiral Lagrangian where the $\rho$ is properly incorporated as a dynamical
degree of freedom, interacting with the nucleons. This, like the pion-nucleon chiral EFT \cite{Prezeau:2003xn}, will enable proper power counting once
$\rho$ meson is also included, such that possible hadronization both into pions and $\rho$ is incorporated in a systematic fashion. 
Such a chiral EFT can then be used for phenomenological analysis since these hadron level contributions 
(local here for the $\rho$ while non-local for pion exchange) have significant impact. 
Beyond this step of lowest order evaluation within such a chiral EFT, lies systematic
calculation of higher order contributions. As mentioned before, neutrinoless double beta decay has been heralded as a harbinger of new physics,
and it is of utmost importance and relevance to have as precise predictions (for a given NME) for the half life to infer the underlying
mechanism of lepton number violation.



%

\end{document}